# Prompt fission neutron spectra of $^{240}$Pu(n, F)

V. M. Maslov[1]
*220025 Minsk, Byelorussia*
E–mail: mvm2386@yandex.ru

Variation of fission neutron spectra of $^{240}$Pu(*sf*) and $^{240}$Pu(*n,F*) for $E_n \lesssim 20$ MeV is predicted. Features of angle-integrated prompt fission neutron spectra (PFNS) of $^{240}$Pu(*n,F*) stem from simultaneous analysis of data for $^{238}$U(*n,F*), $^{239}$Pu(*n,F*) and available data on $^{240}$Pu(*n,F*) PFNS. The data on average energies $\langle E \rangle$ of $^{240}$Pu(*n,F*) PFNS support the approach pursued in case of $^{238}$U(*n,F*) and $^{239}$Pu(*n,F*). Soft influence of exclusive neutron spectra on PFNS is observed in case of $^{240}$Pu(*n,F*) and $^{240}$Pu$(n,nf)^1$ at $E_n$~7÷8 MeV. The largest relative amplitude of exclusive neutron spectra $^{240}$Pu$(n,xnf)^1$ is envisaged at $E_n$~6÷6.25 MeV. PFNS of $^{240}$Pu(*n,F*) are harder than those of $^{238}$U(*n,F*), but softer than PFNS for $^{239}$Pu(*n,F*). The $^{240}$Pu(*n,F*) PFNS shape is rather close to that of $^{239}$Pu(*n,F*), though the contribution of pre-fission neutrons is relatively higher. Exclusive neutron spectra $(n,xnf)^{1,..x}$ are consistent with σ(*n,F*) of $^{237\text{-}240}$Pu(*n,F*), as well as neutron emissive spectra of $^{239}$Pu(*n,xn*) at $E_n$~14 MeV. We predict the $^{240}$Pu$(n,xnf)^{1,..x}$ exclusive pre–fission neutron spectra, exclusive neutron spectra of $^{240}$Pu$(n,xn)^{1,..x}$ reactions, average total kinetic energy TKE of fission fragments and products, partials of average prompt fission neutron number and observed PFNS of $^{240}$Pu(*n,F*).

## INTRODUCTION

Since the recent paper [1] on $^{238}$U(*n,F*), $^{239}$Pu(*n,F*) and $^{240}$Pu(*n,F*) PFNS, became available fragmentary data on PFNS and its $\langle E \rangle$ for 0.8 $\lesssim\varepsilon\lesssim$10 MeV prompt fission neutron energy interval [2]. The PFNS shapes of $^{238}$U(*n,F*) [1, 3] and $^{239}$Pu(*n,F*) [1, 4, 5] reactions are very much different from each other, the PFNS of $^{240}$Pu(*n,F*) lying in-between. It seems the contribution of pre-fission neutrons for $^{240}$Pu(*n,F*) is lower than in case of $^{238}$U(*n,F*), but higher than in case of $^{239}$Pu(*n,F*) and vice versa as regards the prompt fission neutrons emitted from fragments. At $E_n \gtrsim E_{nnf}$, $E_{nnf}$ being (*n,nf*) reaction threshold, pre–fission neutrons influence the partitioning of fission energy between excitation energy and TKE of fission fragments. Analysis of measured data for $^{238}$U(*n,F*) and $^{240}$Pu(*n,F*) allows to demonstrate sensitivities of PFNS shape near (*n,xnf*) reaction thresholds to the exclusive pre–fission neutron spectra. Those for $^{238}$U(*n,F*) PFNS [3] are strongly supported by the measured data [6] with minor exceptions at $E_n$~8÷12 MeV. In that case strong influence of $^{238}$U$(n,nf)^1$ excusive neutron spectra on PFNS at $E_n$~7 MeV and $E_n$~7÷8 MeV is demonstrated [1], while it might be predicted for the $^{240}$Pu(*n,F*) and $^{240}$Pu$(n,nf)^1$. PFNS for $^{239}$Pu(*n,F*) [1, 4, 5] are strongly supported by measured data [7, 8]. Much softer influence of $^{240}$Pu$(n,nf)^1$ excusive neutron spectra on PFNS at $E_n \gtrsim E_{nnf}$ is due to exceptionally high contribution of first chance $^{240}$Pu(*n,f*) fission reaction σ(*n,f*) to observed σ(*n,F*).

## $^{240}$Pu(*n,f*) PROMPT FISSION NEUTRON SPECTRA

The reliability of the modelling of PFNS for neutron-induced fission of $^{239}$Pu might be augmented by analysis of $^{239}$Pu($n_{\text{th}}$,*f*)&$^{240}$Pu(*sf*) data sets. Similar augmentation is possible for a pair of fission reactions $^{241}$Pu($n_{\text{th}}$,*f*)&$^{242}$Pu(*sf*). In reaction $^{239}$Pu($n_{th}$,*f*) neutron yield comes mostly from $J^{\pi} = 0^+$ states, same as in $^{240}$Pu(*sf*) spontaneous fission neutron spectra (SFNS). Comparison of PFNS of $^{239}$Pu($n_{th}$,*f*) and SFNS of $^{240}$Pu(*sf*) [9] in [1] shows that at ε<0.2 MeV PFNS and SFNS of fissioning nuclide $^{240}$Pu depend only weakly on the excitation energy, while at $\varepsilon \gtrsim \langle E \rangle$ the PFNS of $^{239}$Pu($n_{th}$,*f*) is much harder than SFNS. The same happens in case of calculated $^{241}$Pu($n_{\text{th}}$,*f*) PFNS and $^{242}$Pu(*sf*) [9]. For a pair $^{240}$Pu($n_{\text{th}}$,*f*)&$^{241}$Pu(*sf*) such augmentation is hardly possible, though some guidance stems from pair of spectra of $^{240}$Pu($n_{\text{th}}$,*f*)&$^{240}$Pu(*sf*). Major parameters of $^{240}$Pu($n_{th}$,*f*) PFNS modelling α, $\alpha_1$ and $E_F^{pre}$ define the kinetic energy of the fragments at the moment of prompt fission neutron emission [1, 3–5]. Lowering of PFNS $\langle E \rangle$ for $^{240}$Pu($n_{th}$,*f*) relative to $\langle E\rangle$ of $^{239}$Pu($n_{th}$,$f$) is due to binding energies $B_n$ and TKE values differences[1, 4, 5]. Figure 1 shows the comparison of $^{240}$Pu($n_{th}$,$f$) PFNS and SFNS of $^{240}$Pu($sf$) [9], the former being harder at $\varepsilon \gtrsim \langle E\rangle$. The $^{239}$Pu($n$,$f$) PFNS data at $E_n$~1.5 MeV [2] are consistent with hardening of the calculated PFNS with increase of $E_n$.

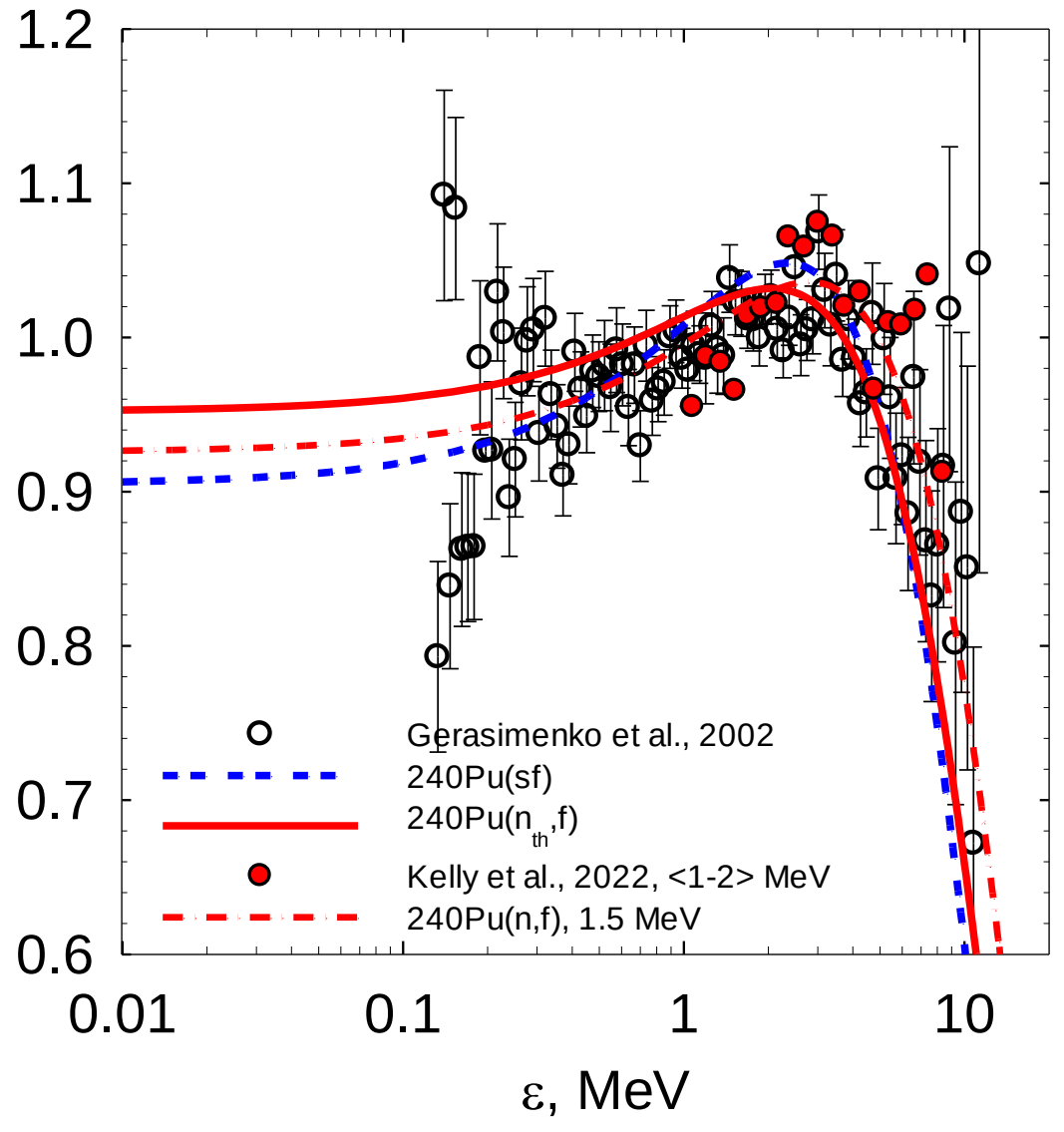

**Fig.1** PFNS of $^{240}$Pu($n_{th}$,$f$) and $^{240}$Pu($n_{th}$,$f$) relative Maxwellian with $\langle E\rangle$=2.0564 MeV.

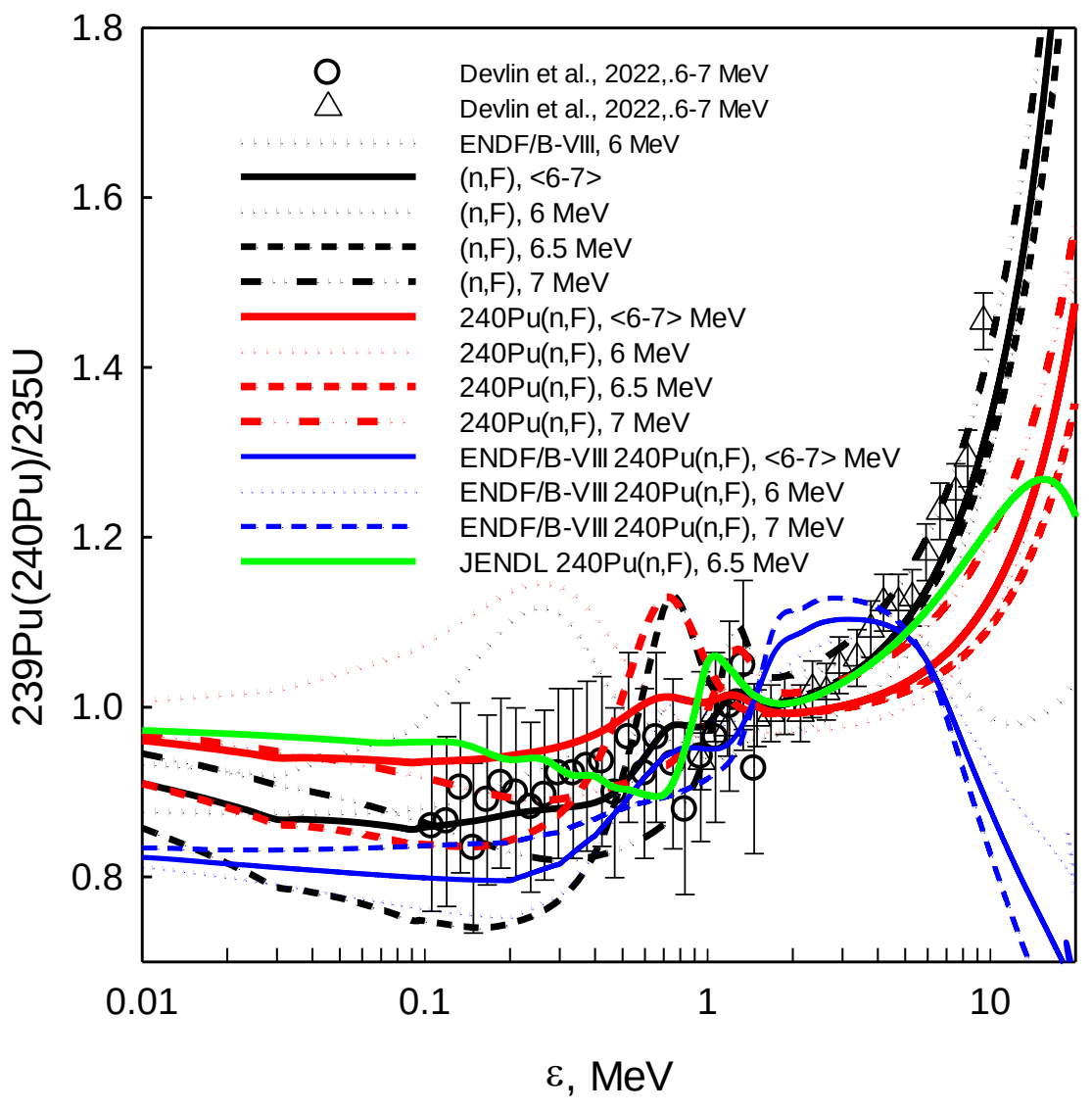

Fig.2 Ratios of PFNS $^{240}$Pu($n$,$F$)/$^{235}$U($n$,$F$) and $^{239}$Pu($n$,$F$)/$^{235}$U($n$,$F$) at $E_n$~6÷7 MeV.

## ($n$,$xnf$) PROMPT FISSION NEUTRON SPECTRA

At $E_n > E_{nnf}$ integral prompt fission neutron spectra $S(\varepsilon, E_n)$ is a superposition of exclusive spectra of pre-fission neutrons, $(n,nf)^1$, $(n,2nf)^{1,2}$, $(n,3nf)^{1,2,3}$ – $\frac{d\sigma^k_{nxnf}(\varepsilon,E_n)}{d\varepsilon}$ ($x$=0, 1, 2, 3; $k$=1,…,$x$), and spectra of prompt fission neutrons, emitted by fission fragments, $S_{A+1-x}(\varepsilon, E_n)$:

$$\begin{aligned}
&S(\varepsilon,E_n)=\tilde{S}_{A+1}(\varepsilon,E_n)+\tilde{S}_{A}(\varepsilon,E_n)+\tilde{S}_{A-1}(\varepsilon,E_n)+\tilde{S}_{A-2}(\varepsilon,E_n)=\\
&\nu_p^{-1}(E_n)\cdot\{\ \nu_{p1}(E_n)\cdot\beta_1(E_n)S_{A+1}(\varepsilon,E_n)+\nu_{p2}(E_n-\langle E_{nnf}\rangle)\beta_2(E_n)S_A(\varepsilon,E_n)+\\
&+\beta_2(E_n)\frac{d\sigma^1_{nnf}(\varepsilon,E_n)}{d\varepsilon}+\nu_{p3}(E_n-B_n^A-\langle E^1_{n2nf}\rangle-\langle E^2_{n2nf}\rangle)\beta_3(E_n)S_{A-1}(\varepsilon,E_n)+\beta_3(E_n)\cdot\\
&\left[\frac{d\sigma^1_{n2nf}(\varepsilon,E_n)}{d\varepsilon}+\frac{d\sigma^2_{n2nf}(\varepsilon,E_n)}{d\varepsilon}\right]+\nu_{p4}(E_n-B_n^A-B_n^{A-1}-\langle E^1_{n3nf}\rangle-\langle E^2_{n3nf}\rangle-\langle E^3_{n3nf}\rangle)\cdot\\
&\beta_4(E_n)S_{A-2}(\varepsilon,E_n)+\beta_4(E_n)\left[\frac{d\sigma^1_{n3nf}(\varepsilon,E_n)}{d\varepsilon}+\frac{d\sigma^2_{n3nf}(\varepsilon,E_n)}{d\varepsilon}+\frac{d\sigma^3_{n2nf}(\varepsilon,E_n)}{d\varepsilon}\right]\ \}.
\end{aligned} \tag{1}$$

In Eq. 1 $\tilde{S}_{A+1-x}(\varepsilon,E_n)$ is contribution of $x$-chance fission to the observed PFNS, $\langle E^k_{nxnf}\rangle$ – average energy of $k$–th neutron of ($n$,$xnf$) reaction with spectrum $\frac{d\sigma^k_{nxn}(\varepsilon,E_n)}{d\varepsilon}$, $k\le x$, spectra $S(\varepsilon,E_n)$, $S_{A+1-x}(\varepsilon,E_n)$ and $\frac{d\sigma^k_{nxn}(\varepsilon,E_n)}{d\varepsilon}$ are normalized to unity, $\beta_x(E_n)=\sigma_{n,xnf}(E_n)/\sigma_{n,F}(E_n)$ – contribution of $x$– th fission chance to the observed fission cross section, $\nu_p(E_n)$ observed average number of prompt fission neutrons, $\nu_{px}(E_{nx})$– average number of prompt fission neutrons, emitted by $^{241-x}$Pu nuclides. Spectra of prompt fission neutrons, emitted from fragments, $S_{A+1-x}(\varepsilon,E_n)$, as proposed in [10], were approximated by the sum of two Watt [11] distributions with different temperatures, the temperature of light fragment being higher. Exclusive neutron spectra $(n,xnf)^{1,..x}$ are calculated simultaneously with σ($n,F$) of $^{237\text{-}240}$Pu($n,F$), as well as neutron emissive spectra of $^{239}$Pu($n,xn$) at ~14 MeV [4, 5].

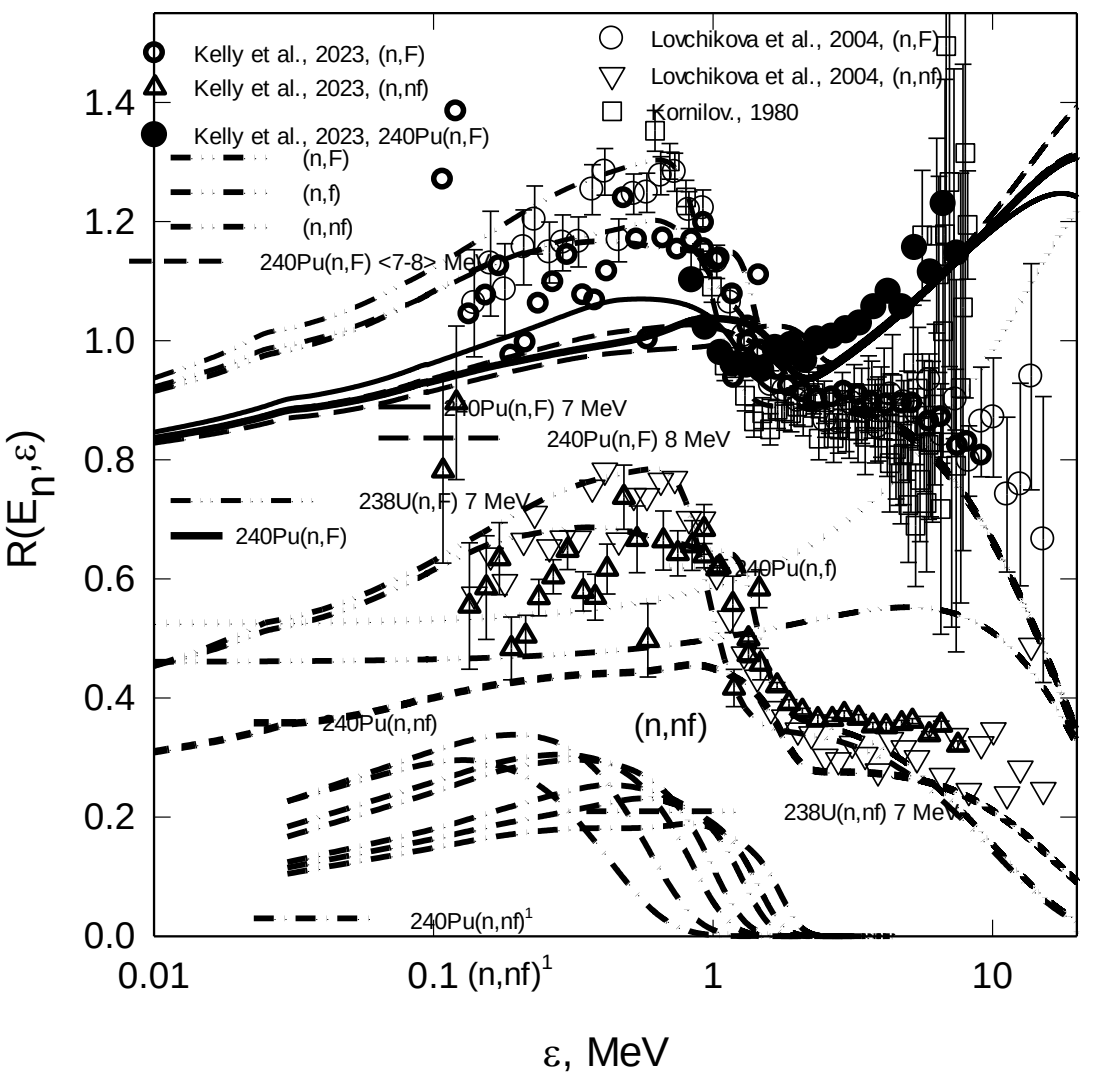

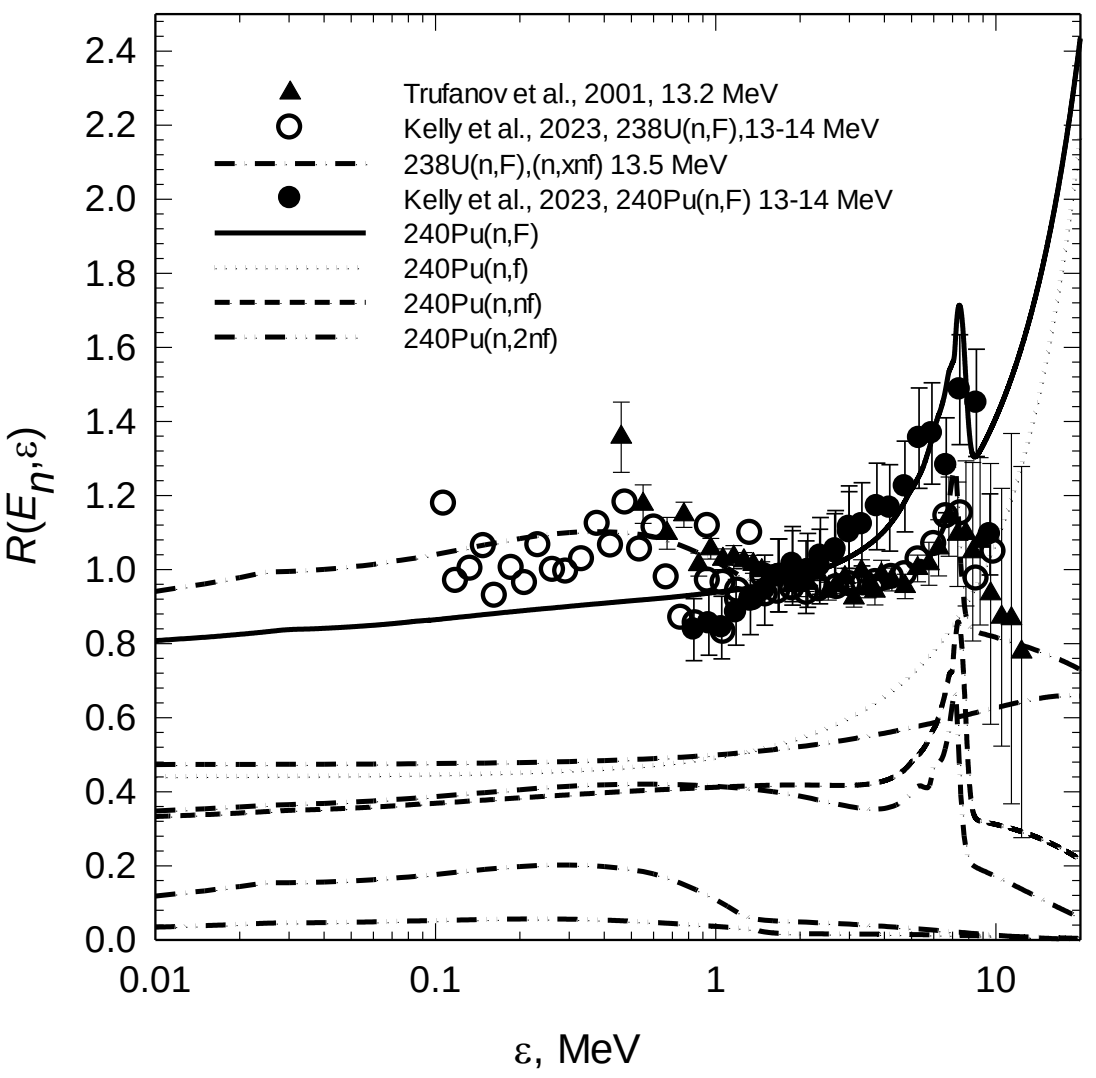

Fig.3 PFNS of $^{240}$Pu($n,F$) and $^{238}$U($n,F$) relative to Maxwellian with $\langle E \rangle$=2.0242 MeV, $E_n$~7÷8 MeV.

Fig.4 PFNS $^{240}$Pu($n,F$) and $^{238}$U($n,F$) relative to Maxwellian with $\langle E \rangle$=2.0415 MeV, $E_n$~13÷14 MeV.

Analysis of prompt fission neutron spectra of $^{238}$U($n,F$) [1, 3, 12, 13] and $^{239}$Pu($n,F$) [1, 4, 5] confirmed strict correlations of a number observed data structures with $(n,xnf)^{1\cdots x}$ pre-fission neutrons. Pre-fission neutron spectra turned out to be quite soft as compared with neutrons emitted by excited fission fragments. The net outcome of that is the decrease of $\langle E \rangle$ in the vicinity of the ($n,xnf$) thresholds. The amplitude of the $\langle E \rangle$ variation is much higher in case of $^{238}$U($n,F$) as compared with $^{239}$Pu($n,F$) [1]. This peculiarity is due to differing emissive fission contributions in $^{239}$Pu($n,F$) and $^{238}$U($n,F$). The increase of the average energies of the exclusive pre-fission neutron spectra is faster in case of $^{239}$U($n,F$) than in case of $^{239}$Pu($n,F$), in consistency with measured data. The differential PFNS are susceptible to systematic errors of different origin. In ratios of PFNS, especially of draft PFNS data, these errors may be cancelled [6, 14]. Figure 2 shows the $^{239}$Pu($n,F$)/$^{235}$U($n,F$) and $^{240}$Pu($n,F$)/$^{235}$U($n,F$) ratios of PFNS for rather wide range of $E_n$~6÷7 MeV. In case of $^{239}$Pu($n,F$)/$^{235}$U($n,F$) ratio the data of [4, 5], when averaged over $E_n$~6÷7 MeV range are compatible with measured data [6, 14]. The averaged $^{240}$Pu($n,F$)/$^{235}$U($n,F$) ratio is very much different from that of $^{239}$Pu($n,F$)/$^{235}$U($n,F$). The $^{239}$Pu($n,F$)/$^{235}$U($n,F$) ratios of differential PFNS at $E_n$~6, 6.5 and 7 MeV fluctuate around averaged values. The $^{240}$Pu($n,F$)/$^{235}$U($n,F$) ratios at $E_n$~6, 6.5 and 7 MeV are also very much different from those of $^{239}$Pu($n,F$)/$^{235}$U($n,F$). That difference is due to shape of exclusive pre-fission ($n,nf$) neutron spectra and $\beta_x(E_n) = \sigma_{n,xnf}(E_n)/\sigma_{n,F}(E_n)$ values. The $^{240}$Pu($n,F$)/$^{235}$U($n,F$) averaged ratios and ratios at $E_n$~6, 6.5 and 7 MeV of ENDF/B-VIII.0 [15] and JENDL-4.0 [16] are much discrepant with our estimates. That might be due at least to the same reasons as in case of our calculated $^{239}$Pu($n,F$)/$^{235}$U($n,F$) and $^{240}$Pu($n,F$)/$^{235}$U($n,F$). In the next energy range of $E_n$~7÷8 [6] the fluctuations of differential PFNS at $E_n$~7, 7.5 and 8 MeV are damped both in case of $^{239}$Pu($n,F$)/$^{235}$U($n,F$) and $^{240}$Pu($n,F$)/$^{235}$U($n,F$) ratios. The $^{240}$Pu($n,F$) PFNS is compared with $^{238}$U($n,F$) PFNS measured [6, 17, 18] and calculated [1, 13] data on Fig. 3. Obviously, the shapes of $^{240}$Pu $(n,nf)^1$ exclusive pre-fission neutron spectra are much different from those of $^{238}$U$(n,nf)^1$. The PFNS of $^{240}$Pu($n,F$) at $E_n$~7, 7.5 and 8 MeV are not very much dependent upon excitation energy, since the threshold of $^{240}$Pu($n,2n$) is lower than that of $^{238}$U($n,2n$).
That influences competition of $(n,nf)^1$ and $(n,2n)^{1,2}$ reactions. At $E_n \gtrsim E_{n2nf}$ integral emission spectrum of $(n,nX)^1$ reaction, $\frac{d^2\sigma^1_{nnx}(\varepsilon,E_n)}{d\varepsilon}$, could be represented as a sum of compound and weakly dependent on neutron emission angle pre-equilibrium components, and phenomenological function, modelling energy and angle dependence of neutron spectra, relevant for the $^{240}$Pu excitations of 1~6 MeV. Angle-averaged $\langle \omega(\theta) \rangle_\theta$ function, $\omega(\theta)$ [5], is approximated as $\langle \omega(\theta) \rangle_\theta \approx \omega(90^o)$, then integral spectrum equals

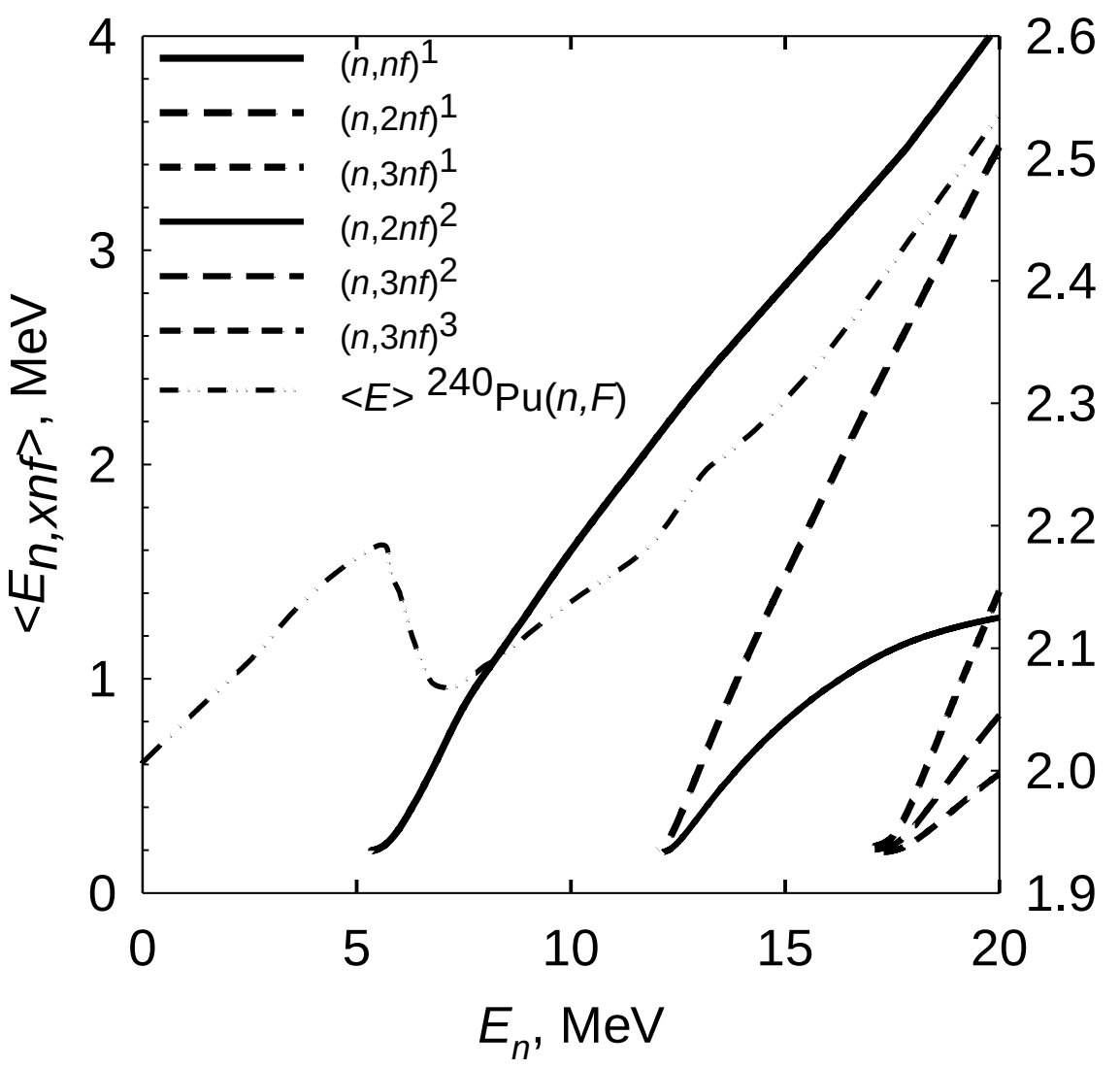

Fig.5 Average energy $\left\langle E^{k}_{nxnf}\right\rangle$ of $^{240}$Pu($n$,$F$) PFNS

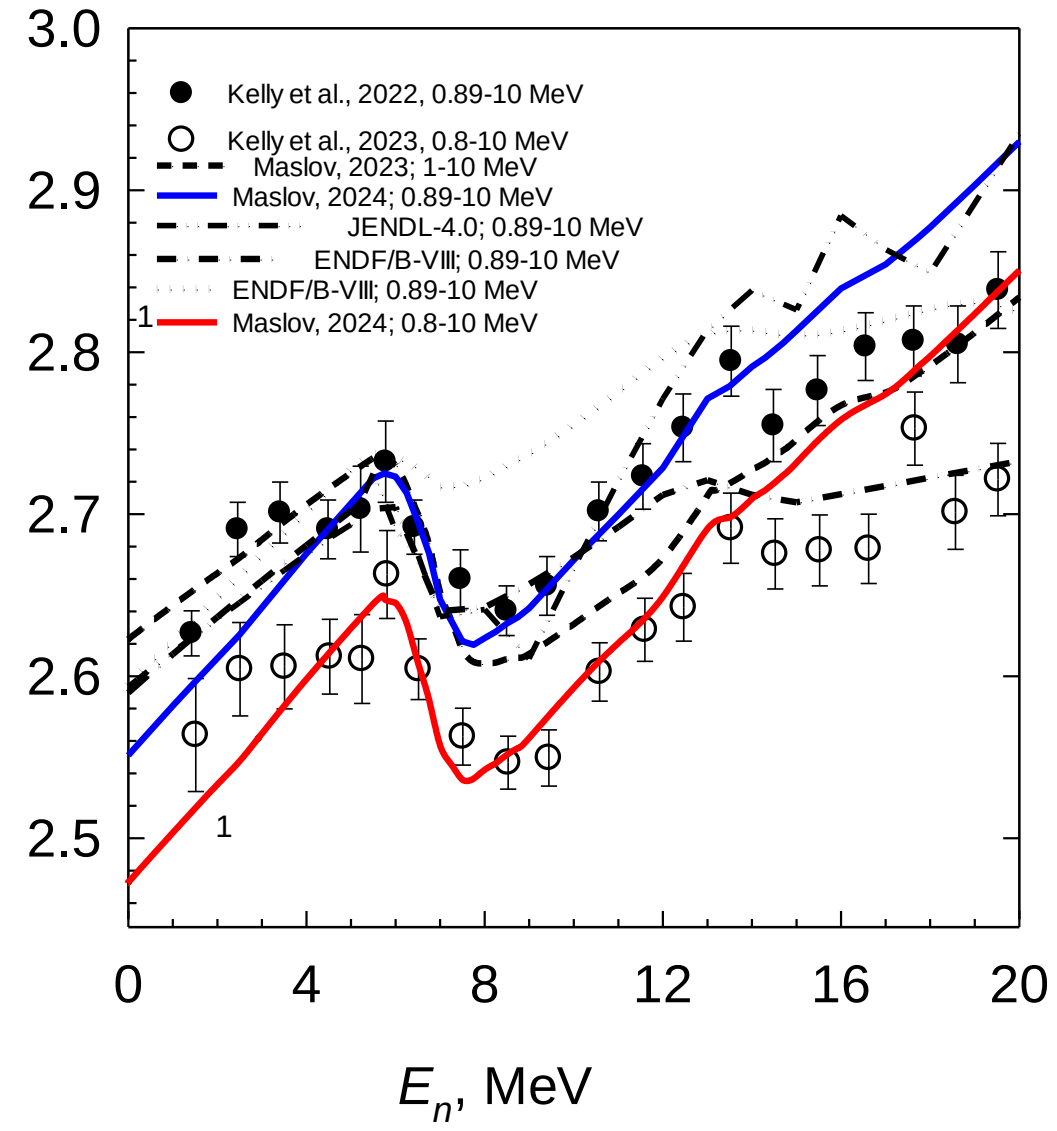

Fig.6 Average energy $\left\langle E\right\rangle$ of $^{240}$Pu($n$,$F$) PFNS

$$\frac{d\sigma^{1}_{nnx}(\varepsilon,E_n)}{d\varepsilon}\approx\frac{d\tilde{\sigma}^{1}_{nnx}(\varepsilon,E_n)}{d\varepsilon}+\sqrt{\frac{\varepsilon}{E_n}}\,\frac{\left\langle\omega(\theta)\right\rangle_{\theta}}{E_n-\varepsilon}\qquad(2)$$

Figure 4 compares calculated and measured PFNS of $^{238}$U($n$,$F$) [6, 19] and $^{240}$Pu($n$,$F$) [2] at $E_n$~13÷14 MeV. The $\tilde{S}_{241-x}(\varepsilon,E_n)$ components of PFNS, shown on Fig. 4 reveal high contribution of $\tilde{S}_{241}(\varepsilon,E_n)$ at ε≳$\left\langle E\right\rangle$. The contribution of $\tilde{S}_{240}(\varepsilon,E_n)$ is lower than respective second-chance contribution of $\tilde{S}_{238}(\varepsilon,E_n)$. The contributions of $\frac{d\sigma^{1}_{n2nf}(\varepsilon,E_n)}{d\varepsilon}$ and $\frac{d\sigma^{2}_{n2nf}(\varepsilon,E_n)}{d\varepsilon}$ to the third-chance fission component $\tilde{S}_{239}(\varepsilon,E_n)$ are much lower than in case of $^{238}$U($n$,2$nf$) reaction, as predicted in [3].

Pre-fission neutrons define PFNS shape of $^{240}$Pu($n$,$F$) at $E_n$~$E_{nnf}$–20 MeV. The variations of observed mean energies $\left\langle E\right\rangle$ for 0≲ε≲30 MeV in the vicinity of ($n$,$xnf$) reaction thresholds, are shown on Fig. 5. The excitation energy residual nuclides, after emission of ($n$,$xnf$) neutrons, is decreased by the binding energy of emitted neutron $B_{nx}$ and its average kinetic energy:

$$U_x=E_n+B_n-\sum_{x,1\le k\le x}(<E^{k}_{nxnf}(\theta)>+B_{nx})\,.\qquad(3)$$

Values $\left\langle E^{k}_{nxnf}\right\rangle$ of exclusive spectra of ($n$,$xnf$)$^{1,..x}$ pre-fission neutrons are shown on Fig. 5.

The amplitude of variations of $\left\langle E\right\rangle$ for $^{240}$Pu($n$,$F$) PFNS for 0.8≲ε≲10 MeV is supported by measured data of [2], preliminary $\left\langle E\right\rangle$ for 0.89≲ε≲10 MeV range are also compatible with present PFNS (see Fig. 6). The estimate [1] of $\left\langle E\right\rangle$ for 1≲ε≲10 MeV exhibits lowering, starting from $E_n$ ≳8 MeV, by ~100 keV. The variations of average PFNS energy due to exclusive spectra of ($n$,$xnf$)$^{1,..x}$ neutrons are the same as predicted in [1], the systematic lowering is due to erroneous $S_{239}(\varepsilon,E_n)$ of $^{238}$U($n$,$f$) reaction. The net effect of these procedures is the adequate approximation of angular distributions of $^{238}$U ($n$,$nX$)$^1$ first neutron inelastic scattering in continuum which corresponds to $U$=1~6 MeV excitations for $E_n$ up to ~20 MeV.

## CONCLUSIONS

Predicted distribution of fission energy of $^{240}$Pu($n$,$F$) reaction between fission fragments kinetic energy, their excitation energy and pre-fission neutrons is compatible with available data on TKE shape of fission fragments and products [1], prompt neutron number, observed PFNS and their average energies.

The estimates of $\left\langle E\right\rangle$ for $^{240}$Pu($n$,$F$) are strongly correlated with PFNS shape, which is not the case in [15, 16]. The influence of exclusive neutron spectra of $^{240}$Pu($n$,$nf$)$^{1}$ and $^{240}$Pu($n$,2$nf$)$^{1,2}$ which they exert on $\langle E \rangle$ in case of $^{240}$Pu($n$,$F$) is much stronger than in case of $^{239}$Pu($n$,$F$), but weaker than in case of $^{238}$U($n$,$F$) PFNS. The correlations of $\langle E \rangle$ variations for $^{240}$Pu($n$,$F$) in the vicinity of $^{240}$Pu($n$,$nf$) and $^{240}$Pu($n$,2$nf$) thresholds, with PFNS shape and values of $\beta_x(E_n) = \sigma_{n,xnf}/\sigma_{n,F}$ , exclusive neutron spectra ($n$,$xnf$)$^{1\ldots x}$ and calculated and observed TKE are established. The abrupt lowering of measured $\langle E \rangle$ of $^{240}$Pu($n$,$F$) PFNS at $E_n \gtrsim E_{n2nf}$ is incompatible with rather mild contribution of $^{240}$Pu($n$,2$nf$) to calculated PFNS.